\input harvmac
\input epsf
\noblackbox
\def\lsw{\lambda_{\rm SW}}
\def\sn{{\rm sn}}
\def\cn{{\rm cn}}
\def\dn{{\rm dn}}
\def\Ze{{\rm Z}}
\def\bfone{\relax{\rm 1\kern-.35em 1}}
\def\inbar{\vrule height1.5ex width.4pt depth0pt}

\def\IC{\relax\,\hbox{$\inbar\kern-.3em{\mss C}$}}
\def\ID{\relax{\rm I\kern-.18em D}}
\def\IF{\relax{\rm I\kern-.18em F}}
\def\IH{\relax{\rm I\kern-.18em H}}
\def\II{\relax{\rm I\kern-.17em I}}
\def\IN{\relax{\rm I\kern-.18em N}}
\def\IP{\relax{\rm I\kern-.18em P}}
\def\IQ{\relax\,\hbox{$\inbar\kern-.3em{\rm Q}$}}
\def\us#1{\underline{#1}}
\def\IR{\relax{\rm I\kern-.18em R}}
\font\cmss=cmss10 \font\cmsss=cmss10 at 7pt
\def\ZZ{\relax\ifmmode\mathchoice
{\hbox{\cmss Z\kern-.4em Z}}{\hbox{\cmss Z\kern-.4em Z}}
{\lower.9pt\hbox{\cmsss Z\kern-.4em Z}}
{\lower1.2pt\hbox{\cmsss Z\kern-.4em Z}}\else{\cmss Z\kern-.4em
Z}\fi}
\def\a{\alpha}

\def\nup#1({Nucl.\ Phys.\ $\us {B#1}$\ (}
\def\plt#1({Phys.\ Lett.\ $\us  {B#1}$\ (}
\def\cmp#1({Comm.\ Math.\ Phys.\ $\us  {#1}$\ (}
\def\prp#1({Phys.\ Rep.\ $\us  {#1}$\ (}
\def\prl#1({Phys.\ Rev.\ Lett.\ $\us  {#1}$\ (}
\def\prv#1({Phys.\ Rev.\ $\us  {#1}$\ (}
\def\mpl#1({Mod.\ Phys.\ Let.\ $\us  {A#1}$\ (}
\def\ijmp#1({Int.\ J.\ Mod.\ Phys.\ $\us{A#1}$\ (}
\def\jag#1({Jour.\ Alg.\ Geom.\ $\us {#1}$\ (}
\def\tit#1|{{\it #1},\ }

\def\Coe#1.#2.{{#1\over #2}}

\def\coe#1.#2.{\relax{\textstyle {#1 \over #2}}\displaystyle}

\def\br{\hfill\break}
%
%
\lref\SW{N.\ Seiberg and E.\ Witten, \nup426(1994) 19,
hep-th/9407087; \nup431(1994) 484, hep-th/9408099.}
\lref\KKLMV{S.\ Kachru, A.\ Klemm, W.\ Lerche, P.\ Mayr and
C.\ Vafa, \nup{459} (1996) 537, hep-th/9508155.}
\lref\KLMVW{A.\ Klemm, W.\ Lerche, P.\ Mayr, C.\ Vafa and
N.P.\ Warner, \nup{477} (1996) 746, hep-th/9604034.}
\lref\witcom{E.\ Witten, {\it Some Comments on String Dynamics},
in ``Strings '95: Future Perspectives in String Theory'',
Editors: I.~Bars, P.~Bouwknegt, J.~Minahan, D.~Nemeschansky, K.~Pilch,
H.~Saleur and N.P.~Warner, Proceedings of USC conference, Los Angeles,
March 13-18, 1995, World Scientific, Singapore, (1996);
hep-th/9507121.}
\lref\strop{A.\ Strominger, \plt{383} (1996) 44, hep-th/9512059.}
\lref\selfd{ O.\ Ganor and A.\ Hanany, \nup{474} (1996) 122,
hep-th/9602120;\br N.\ Seiberg and E.\ Witten, \nup{471} (1996) 121,
hep-th/9603003;\br M.\ Duff, H.\ Lu and C.N.\ Pope, \plt{378}
(1996) 101, hep-th/9603037.}
\lref\ABSS{A.\ Brandhuber and S.\ Stieberger, 
{\it Self-dual Strings and Stability of BPS States in $N=2$ $SU(2)$ 
Gauge Theories,} hep-th/9610053.}
\lref\WLer{W.\ Lerche,
{\it Introduction to Seiberg-Witten Theory and its Stringy Origin},
hep-th/9611190.}
\lref\JRabin{J.\ Rabin, in preparation.}
\lref\GHL{C.\ G\'omez, R.\ Hern\'andez and E.\ L\'opez,
{\it $K3$-Fibrations and Softly Broken $N=4$ Supersymmetric Gauge 
Theories}, hep-th/9608104.}
\lref\russ{A.\ Gorskii, I.\ Krichever, A.\ Marshakov, A.\ Mironov
and  A.\ Morozov, \plt{355} (1995) 466, hep-th/9505035.}
\lref\MW{E.\ Martinec and N.P.\ Warner, \nup{459} (1996) 97,
hep-th/9509161}
\lref\otherinteg{T.\ Nakatsu and K.\ Takasaki, \mpl{11}
(1996) 157, hep-th/9509162.}
\lref\RDEW{R.\ Donagi and E.\ Witten, \nup{460} (1996) 299,
hep-th/9510101.}
\lref\HIAM{H.\ Itoyama and A.\ Morozov, \nup{477} (1996) 855,
hep-th/9511126; and ITEP-M6-95, hep-th/9512161.}
\lref\EMa{E.\ Martinec, \plt{367}, (1996) 91, hep-th/9510294}
\lref\WW{E.T.\ Whittaker and G.N.\ Watson, {\it A Course in Modern
Analysis,} 4th Edition, C.U.P.\ (1927).}
\lref\ABFF{A.\ Bilal and F.\ Ferrari, \nup{480} (1996) 589-622,
hep-th/9605101.}
\lref\FFer{F.\ Ferrari, 
{\it Duality and BPS spectra in $N=2$ supersymmetric QCD},
Nucl.\ Phys.\ B proc.\ sup., hep-th/9611012.}
\lref\ABilal{A.\ Bilal, \nup{469} (1996) 387, hep-th/9602082; 
\nup{480} (1996) 589, hep-th/9605101; {\it Discontinuous BPS 
Spectra in $N=2$ Susy QCD}, hep-th/9606192.}
\lref\ASen{A.~Sen, \plt{329} (1994) 217,  hep-th/9402032.}
\lref\MPor{M.~Porrati, \plt{377} (1996) 67,  hep-th/9505187.}
\lref\FMVW{P.~Fendley, S.D.~Mathur, C.~Vafa and N.P.~Warner,
\plt{243} (1990) 257.}
%

\Title{\vbox{
\hbox{USC/97-001}
\hbox{\tt hep-th/9702012}
}}{BPS Geodesics in $N=2$ Supersymmetric Yang-Mills Theory}

\bigskip
\centerline{J.~Schulze}
\bigskip
\centerline{and}
\bigskip
\centerline{N.P.~Warner}
\bigskip
\centerline{{\it Physics Department, U.S.C.}}
\centerline{{\it University Park, Los Angeles, CA 90089}}

\vskip .3in

We introduce some techniques for making a more global analysis of the
existence of geodesics on a Seiberg-Witten Riemann surface with metric
$ds^2 = |\lambda_{SW}|^2$.  Because the existence of such geodesics
implies the existence of BPS states in $N=2$ supersymmetric Yang-Mills
theory, one can use these methods to study the BPS spectrum in various
phases of the Yang-Mills theory.  By way of illustration, we show how,
using our new methods, one can easily recover the known results for
the $N=2$ supersymmetric $SU(2)$ pure gauge theory, and we show in
detail how it also works for the $N=2$, $SU(2)$ theory coupled to a
massive adjoint matter multiplet.

\vskip .3in


\Date{\sl February, 1997}

%
\parskip=4pt plus 15pt minus 1pt
\baselineskip=15pt plus 2pt minus 1pt
%
\newsec{Introduction}

One of the many remarkable features of string duality is that one has
been able to use it to extract new statements about the strong
coupling regime of supersymmetric field theories.  In particular, one
can re-derive the quantum effective actions of Seiberg and Witten \SW\
from the classical effective actions of type II string
compactifications on K3-fibrations \refs{\KKLMV,\KLMVW}.  In the IIB
theory, the Yang-Mills BPS states come from $3$-branes, and when these
are wrapped around $2$-cycles in the fibration, the result can be
reinterpreted as a six-dimensional self-dual string
\refs{\witcom,\strop,\selfd} compactified to four dimensions on the
Seiberg-Witten Riemann surface, $\Sigma$ \KKLMV.  The BPS states of
Yang-Mills theory then become minimum energy winding configurations of
the self-dual string on $\Sigma$, where the (local) tension in the
string is given by the Seiberg-Witten differential, $\lsw$.

In \SW, the differential $\lsw$ was an object whose period integrals
gave the central charges, and hence the masses of BPS states.  Since
one integrated it around cycles, one was only interested in its
cohomology class -- one was free to add the derivative of any
meromorphic function.  The stringy approach led to a sharper statement
about the BPS states: it is only a specific {\it local} form of $\lsw$
that has the interpretation of a string tension on $\Sigma$, and the
existence of BPS states with given monopole and electric charges
$(\vec g, \vec q)$, is equivalent to the existence of a geodesic with
these winding numbers on $\Sigma$ with the metric
\eqn\metric{ds^2 ~=~  \left|\lsw\right|^2 \ .}
Thus a statement about the stability and existence of strong quantum
BPS states is reduced to a classical computation.

There have been several attempts to use the foregoing as a tool to
probe the BPS structure of the theory \refs{\KLMVW,\ABSS,\WLer,
\JRabin}, but existence of geodesics can be subtle to establish,
particularly if one proceeds numerically.  One would also like to see
precisely what happens as one crosses inside a curve of marginal
stability, where some of the BPS geodesics must ``cease to exist''. 
A proper understanding of this has so far proven elusive.

Our purpose in this paper is to introduce some analytical tools by
which these issues can be addressed.  The key idea is to look for
``geodesic horizons''.  These are maximal, closed geodesics that
surround poles in $\lsw$, and have the property that once crossed by a
BPS geodesic, they can never be recrossed (hence the name ``horizon'')
by a BPS geodesic of finite energy.  We find that, at least in the
$SU(2)$ pure gauge theory, and in the $SU(2)$ gauge theory with
adjoint matter, these geodesic horizons confine the winding states on
$\Sigma$ in a manner that provides a simple geometric understanding of
the BPS spectrum.  Since these geodesic horizons are straighforward to
characterize and their behaviour at curves of marginal stability can
be easily addressed, we anticipate that they will provide a valuable
technique in using geodesic methods to analyze the BPS spectrum of
more complex models than the ones discussed here.

We start in the next section by introducing geodesic horizons and
deriving some of their properties.  We then describe an illustrative,
but unphysical ``toy model''.  In section~3 we obtain some relatively
simple analytic expressions for the indefinite integrals of the
Seiberg-Witten differential in the softly broken $N=4$, $SU(2)$ gauge
theory, and in its $N=2$ supersymmetric pure gauge limit.  In
section~4 we analyze the BPS spectra in various phases of these gauge
theories by using geodesic horizons, and the ``shadows'' cast by such
horizons.

\newsec{BPS geodesics and horizons}

There are a few implicit issues in the geodesic characterization of
BPS states, and these need to be brought into the open.  First, the
geodesic characterization has only been carefully established for pure
gauge and for the $N=4$ supersymmetric model.  While it is almost
certainly true in greater generality, one still needs the stringy
derivation in order to get the proper local form of $\lsw$ so as to
properly describe the local string tension.  Such a stringy derivation
has been obtained for the softly broken $N=4$ supersymmetric, $SU(2)$
gauge theory \GHL. More generally one appeals to the underlying
integrable hierarchy to posit the proper local form of $\lsw$.  That
is, one finds that the indefinite integral of the differential,
$\lsw$, selected by the string theory is the Hamilton--Jacobi function
of the integrable hierarchy that underlies the construction of the
effective action \refs{\russ,\MW,\otherinteg,\RDEW}. It is natural to
assume that this remains true in general.

The second issue concerns what constitutes a BPS geodesic.  The
geometrical origin from $3$-branes in ten dimensions implies that
states with magnetic charge must begin and end at ``branch points of
the fibration'' \KLMVW, whereas purely electric states merely wind
with no specific base point.  From the point of view of the Riemann
surface, the branch point is a mere coordinate artefact -- the
invariant statement of this comes from the fact that $\lsw$ has zeroes
at the branch points of the fibration.  We therefore take BPS
geodesics to be those that begin and end at the zeroes of $\lsw$.

Given a set of winding numbers $(\vec g,\vec q)$ for a BPS state there
is {\it always} a corresponding (minimum length) geodesic: one
considers all curves with the same fixed end-points and the same
winding numbers, and then minimizes the length within this homotopy
class.  This geodesic is then either {\it reducible} or {\it
irreducible}.  A geodesic will be called reducible if it is the
concatenation of two other BPS geodesic curves.  It is thus reducible
if it runs into a zero of $\lsw$ at an intermediate point along its
length.  If the shortest curve in the proper homotopy class is
reducible, then the corresponding BPS state is the sum of two others,
and there is no fundamental Yang-Mills BPS state with these quantum
numbers.  The BPS spectrum is thus characterized by irreducible BPS
geodesics, and transitions in the spectrum must correspond to
irreducible geodesics becoming reducible.

The geodesic equation for \metric\ has a trivial implicit solution
\eqn\inteqn{w ~\equiv~\int_{z_0}^z \lsw (z) ~=~
\alpha~t \, ,}
where $z_0$ is the starting point, $t$ is the parameter and $\alpha$
is a constant of integration.  The solution is unique provided that
the tangent vector is continuous, and the only way the tangent can
fail to be continuous is if the curve runs into a zero of $\lsw$ --
{\it i.e.} if the geodesic is irreducible.  Therefore the irreducible
geodesics can be obtained by solving the initial value problem having
used the winding numbers to fix the constant, $\alpha$.  That is, if
$\vec a$ and $\vec a_D$ are the periods of $\lsw$, then the
irreducible geodesic with charges $(\vec g,\vec q)$, if it exists,
must be obtained by taking $\alpha = \vec q \cdot \vec a + \vec g
\cdot \vec a_D$ and letting $t$ run from $0$ to $1$.  If this method
produces a curve that does not terminate (at $t=1$) at a zero of
$\lsw$, then the BPS geodesic with winding numbers $(\vec g,\vec q)$
must be reducible.

The latter method is effective, and was used in \refs{\KLMVW,\ABSS},
but it is somewhat implicit.  To see transitions in the BPS spectrum
it is easier to think in terms of minimizing the lengths of curves
within a homotopy class, and finding that these curves move towards
another zero of $\lsw$ as one approaches a curve of marginal
stability.

Finally, we note  that since the geodesic equation is solved by
\inteqn, it follows that the length of a geodesic, $\Gamma$, is given
by:
\eqn\sumformula{L(\Gamma)  ~=~ \sum_{{\rm irr.\  segs.}\
\gamma}~L(\gamma) ~=~ \sum_{{\rm irr.\  segs.} \  \gamma}~\bigg|~
\int_\gamma~\lsw (z) ~\bigg|\ .}

\subsec{Geodesic horizons}

We wish to show how the existence of certain stable BPS states
precludes the existence of others.  To see this most clearly we will
henceforth work in the covering space, $\widetilde \Sigma$, of the
Riemann surface, $\Sigma$, lifting the metric and curves in the
obvious manner.  In particular we will now take $\lsw(z)$ and the
integral \inteqn\ on $\widetilde \Sigma$.  We will also be interested
in the intersections of various geodesic curves, and by this we will
mean intersections on $\widetilde \Sigma$.  The idea is to see how BPS
geodesics partition the covering space, and this is all a consequence
of three simple facts: (i) geodesics are straight lines in the
$w$-plane (where $w$ is defined in \inteqn), (ii) geodesics
representing BPS states of finite energy are {\it finite} line
segments in the $w$-plane , and (iii) finite line segments (in the
$w$-plane) that intersect more than once must coincide along a common
segment.  An irreducible BPS geodesic must consist of a {\it single}
line segment in the $w$-plane, beginning and ending at zeroes of
$dw/dz= \lsw$, but not encountering any other zeroes of $\lsw$ along
its length.  A reducible geodesic may still be a single line segment,
but generically will be a collection of such segments, and might
involve traversing the same segment twice.

One is thus tempted to conclude that a pair of irreducible geodesics
can only intersect once, but there is a minor subtlety in (iii).
Suppose two geodesics on $\widetilde \Sigma$ intersect at two points,
$P_1$ and $P_2$, then either (a) the two geodesics coincide between
$P_1$ and $P_2$ on $\widetilde \Sigma$, or (b) the geodesics lie on
different branches of $z(w)$, and these branches meet at $P_1$ and
$P_2$.  In the latter instance, $P_1$ and $P_2$ must either be poles
or zeroes of $\lsw$.  Thus if two irreducible geodesics intersect more
than once then they are either identical or they have common
endpoints.

We will say that a closed curve, $\Gamma$, has the {\it horizon
property} if any finite irreducible geodesic that crosses $\Gamma$ can
never recross it (or even return to it).  We will call a closed curve
a {\it geodesic horizon} if a) it is a closed geodesic that has the
horizon property, and b) surrounds a region that contains no zeroes of
$\lsw$.  We will, however, allow geodesic horizons to pass through
zeroes of $\lsw$, and we further define a geodesic horizon to be
reducible if it passes through two or more distinct zeroes of $\lsw$.
The irreducible components of a geodesic horizon are finite line
segments in the $w$-plane.  (In some circumstances there will be
infinitely many concentric geodesic horizons, and so we will
subsequently refine this definition to only mean the outermost, or
maximal such horizon.)

The whole point of excluding zeroes from the interior is that it
ensures that any irreducible geodesic that crosses a geodesic horizon
can never meet a zero, and thus cannot represent a fundamental BPS
state.

We now show that every pole, $z_0$, of $\lsw$ is surrounded by at
least one geodesic horizon.  Consider the family of simple, closed
curves (in $\widetilde \Sigma$) that satisfy the following:
\item{(i)} The only pole of $\lsw$ that they contain is $z_0$
itself.
\item{(ii)} They do not surround any zeroes of $\lsw$
(but can pass through such zeroes).

Within this homotopy class there is at least one curve of minimum
length: A candidate is any geodesic polygon connecting zeroes of
$\lsw$, and surrounding the pole.  One might find a shorter curve
inside the polygon, but the curve of minimal length cannot be trivial
since the length of the curves becomes infinite as they approach the
pole.  These closed curves of minimum length are geodesics, and the
minimality of their length means that they must be simple ({\it i.e.}
not self-crossing).

We now show that any such closed curve, $\Gamma$, has the horizon
property.  If $\Gamma$ is irreducible then the horizon property is
almost obvious.  The only finite irreducible geodesic (apart from
$\Gamma$ itself) that can meet $\Gamma$ more than once is a geodesic
that meets $\Gamma$ only at the (single) zero of $\lsw$ that lies on
$\Gamma$.  Such a geodesic thus cannot be irreducible and
cross~$\Gamma$.

Suppose that $\Gamma$ is reducible and that an irreducible geodesic,
$\gamma$ meets $\Gamma$ at two points, $P_1$ and $P_2$.  Further
suppose the $\gamma$ actually crosses $\Gamma$ at one of these points,
$P_1$.  It follows that $P_1$ cannot be a zero of $\lsw$, since this
would violate the irreducibility of $\gamma$.  Let $\Gamma_1$ and
$\Gamma_2$ be segments of $\Gamma$ between $P_1$ and $P_2$, and let
$\gamma_0$ be the segment of $\gamma$ between $P_1$ and $P_2$.  Since
$w(\gamma_0)$ is a single line segment between $w(P_1)$ and $w(P_2)$
in the $w$-plane, it follows that $L(\gamma_0) \le L(\Gamma_i)$, $i =
1,2$, with equality if and only if $w(\Gamma_i) = w(\gamma_0)$.  Since
$P_1$ is not a zero of $\lsw$, the latter equality would imply that
that $\Gamma_i = \gamma_0$.  Now observe that both $\Gamma_1 \cup
\gamma_0$ and $\Gamma_2 \cup \gamma_0$ are closed curves, neither
contains a zero of $\lsw$\foot{This follows because $\Gamma$ must be
simple.}, and one of them contains the pole.  Suppose that it is the
former.  Minimality of $\Gamma$ then requires $L(\Gamma_2) \le
L(\gamma_0)$, and hence there must be equality.  From the comment
above, we therefore conclude that $\gamma_0 = \Gamma_2$, and hence
$\gamma$ cannot cross into the interior of $\Gamma$.

Thus {\it any} irreducible closed geodesic around a pole has the
horizon property, but a reducible closed geodesic crucially needs to
be the geodesic of globally minimal length in order to have the
horizon property.

Consider now a closed geodesic $\Gamma$ of minimum length that
surrounds several poles.  The foregoing argument fails, but in an
interesting way.  If the geodesic $\gamma$ goes between the poles so
that the total residue in both $\Gamma_1 \cup \gamma_0$ and $\Gamma_2
\cup \gamma_0$ is non-zero, then $\gamma$ can recross $\Gamma$.  If
however, one of the two residues is zero, then it follows from
\sumformula\ that $L(\gamma_0) \le L(\Gamma_i)$, and the argument
still goes through.  In terms of BPS states this means that a BPS
geodesic can emerge from $\Gamma$, provided that the BPS state picks
up a hypermultiplet charge from the interior that is different from
the total hypermultiplet charge enclosed by $\Gamma$.  We will refer
to closed curves like $\Gamma$ as selective horizons.

We now need to refine the definition of irreducible geodesic horizons
since there can be infinitely many of them (see the example below).
If $\gamma_1$ and $\gamma_2$ are two irreducible geodesic horizons
around the same pole, then one geodesic must lie inside the other.
(They cannot cross since they are closed, and would thus have to cross
twice.)  To make the definition more useful, if there is more than one
irreducible horizon around a pole then take the geodesic horizon to be
the {\it maximal} one, that is, the one that is contained in no other
such horizon.  Again, we exclude the possibility of zeroes of $\lsw$
from the interior of the region surrounded horizon.

By definition, a reducible horizon must encounter at least two zeroes
of $\lsw$ upon its length.  It may thus be thought of as a sum of BPS
states whose net electric and magnetic charges are zero.  We now show
that the (maximal) irreducible geodesic horizons must meet one zero of
$\lsw$.

First observe that an irreducible horizon can only surround a pole
with a non-zero residue, and conversely, the horizon around a pole
with no residue must necessarily be reducible.  This is a trivial
consequence of \sumformula\ and residue calculus.  Let the residue of
the pole be $m$, then closed irreducible geodesics around the pole are
given by the straight lines between some point $w_0$ and the point
$w_0 + 2\pi i m$.  Let $\gamma$ be the ``outermost'' such closed
geodesic, and let $w_1$ be a point on it.  Suppose that $\lsw$ is
non-zero everywhere on $\gamma$, then we can locally invert to get
$z(w)$ around $\gamma$.  Look at all straight lines, $S_\epsilon$,
running from $w_1 + \epsilon$ to $w_1 + \epsilon + 2 \pi i m$, and
consider preimages, $\gamma_\epsilon$, in the $z$-plane of
$S_\epsilon$.  Since $\gamma$ is maximal, then no matter how small one
chooses $|\epsilon|$ there must always be some $\gamma_\epsilon$ that
is an {\it open} curve.  Let $z_1$ be the limit point where the
$\gamma_\epsilon$'s first open out.  By considering the integral along
$\gamma_\epsilon$, and upon a small segment $\delta_z$ that closes
$\gamma_\epsilon$ to a loop around the pole, one easily sees that
$\lsw = dw/dz$ must vanish at $z_1$.

To summarize, all geodesic horizons must be closed geodesics loops
that run through at least one of the zeroes of $\lsw$.  They must
consist of irreducible geodesic segments, which can be thought of as
BPS states.  Irreducible horizons can only surround poles with
residues, and horizons around poles with no residue must necessarily
be reducible.  Most imporantantly, any irreducible geodesic that
crosses a horizon can never recross the horizon, and so BPS geodesics,
fundamental or composite, can only exist if they lie outside, or
tangent to, such horizons.  Thus there are ``BPS horizon states''
whose existence and behaviour determines the existence and behaviour
of all the other BPS states.

\subsec{A toy example}

To illustrate these ideas we consider an unphysical example with many
of the features of important physical examples.

Consider the conformal mapping:
\eqn\wtoy{ w ~=~ z  ~+~ {1 \over 2}~  \log \bigg({z - 1 \over
z+1} \bigg) ~-~ { i\pi \over 2} \ .}
The corresponding differential is $\lambda = {z^2 \over z^2 - 1}$ has
a double zero at $z=0$, and two simple poles with residues $\pm 1/2$.
Since $w$ has a triple zero at $z=0$, straight lines in the $w$-plane
through $w=0$ turn a $60^\circ$ corner at $z=0$.  The right
half-$z$-plane maps onto the combination of the right half-$w$-plane
and the strip $\{w: Re(w) < 0, -\pi \le Im(w) < 0 \}$ (see Fig.~1).
The top and bottom of strip may be periodically identified, making the
cylinder that can be associated with the conformal map $z \to
\log(z)$.  The left half-$z$-plane maps in the reflected manner: to
the left half-$w$-plane and the strip, or cylinder $\{w: Re(w) > 0,
-\pi \le Im(w) < 0 \}$.  The two patches are glued together along two
parts of the imaginary $w$-axis: $Im(w) >0$ and $Im(w) < -\pi$.  The
points $w = 0$ and $w = -i\pi$ are to be identified.  As a result,
circling by $2 \pi$ around $z = 0$ results in a circle of $6\pi$ in
the $w$-plane.

The $w$-plane thus looks like the $z$-plane but with two semi-infinite
cylinders sewn into it: one on each side of the line $Re(w) = 0, -\pi
< Im(w) < 0$.  Straight lines parallel to the imaginary $w$-axis, and
lying in the strips are closed geodesic loops around the simple poles
in the $z$-plane.  The geodesic horizons are both mapped to the
straight line in the $w$-plane across the necks of the ``cylinders,''
{\it i.e.} running along $Re(w) = 0, - \pi \le Im(w) \le 0$ (see
Fig.~1.).  Clearly, any straight line in the $w$-plane that crosses
one of these horizons will spiral down the cylinder, never to return.

\goodbreak\midinsert
\vskip .5cm
\centerline{ {{\epsfxsize 5cm\epsfbox{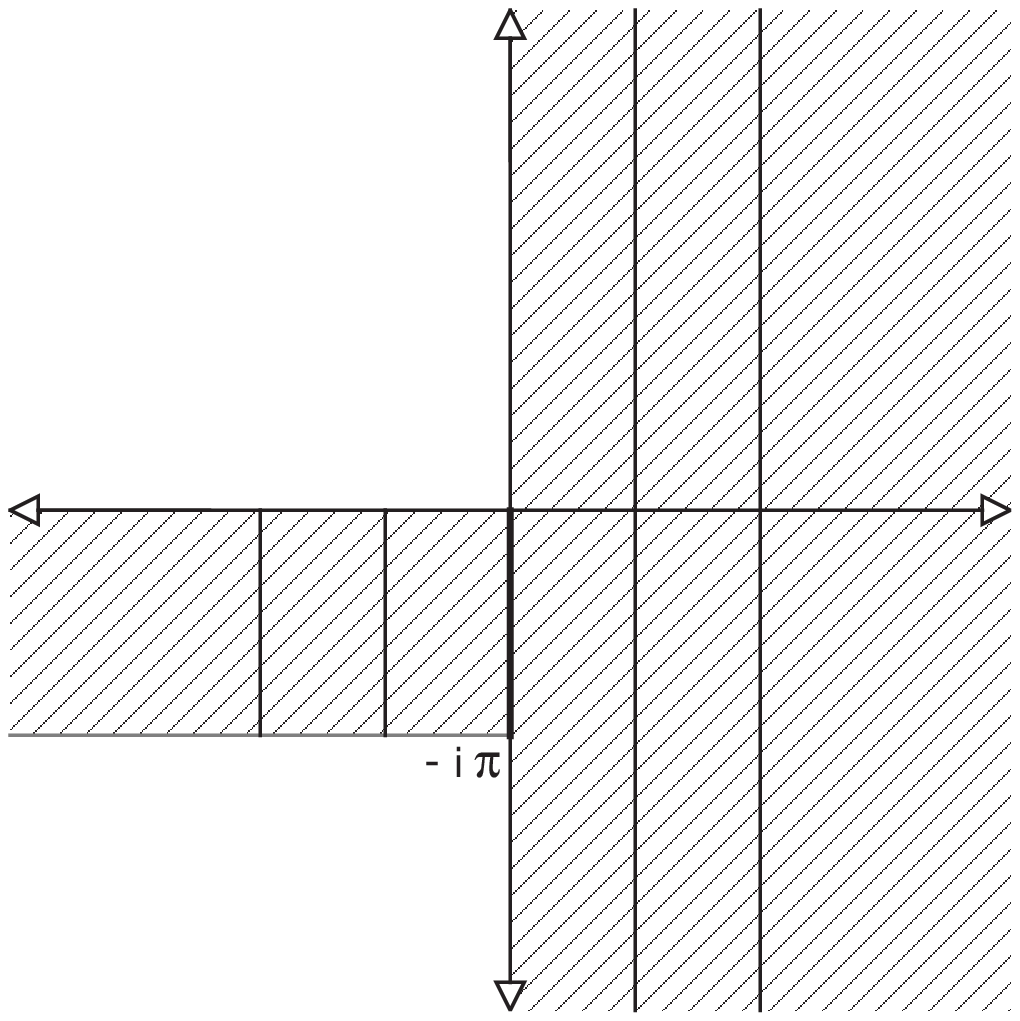}}
\hskip 2cm {\epsfxsize 5cm\epsfbox{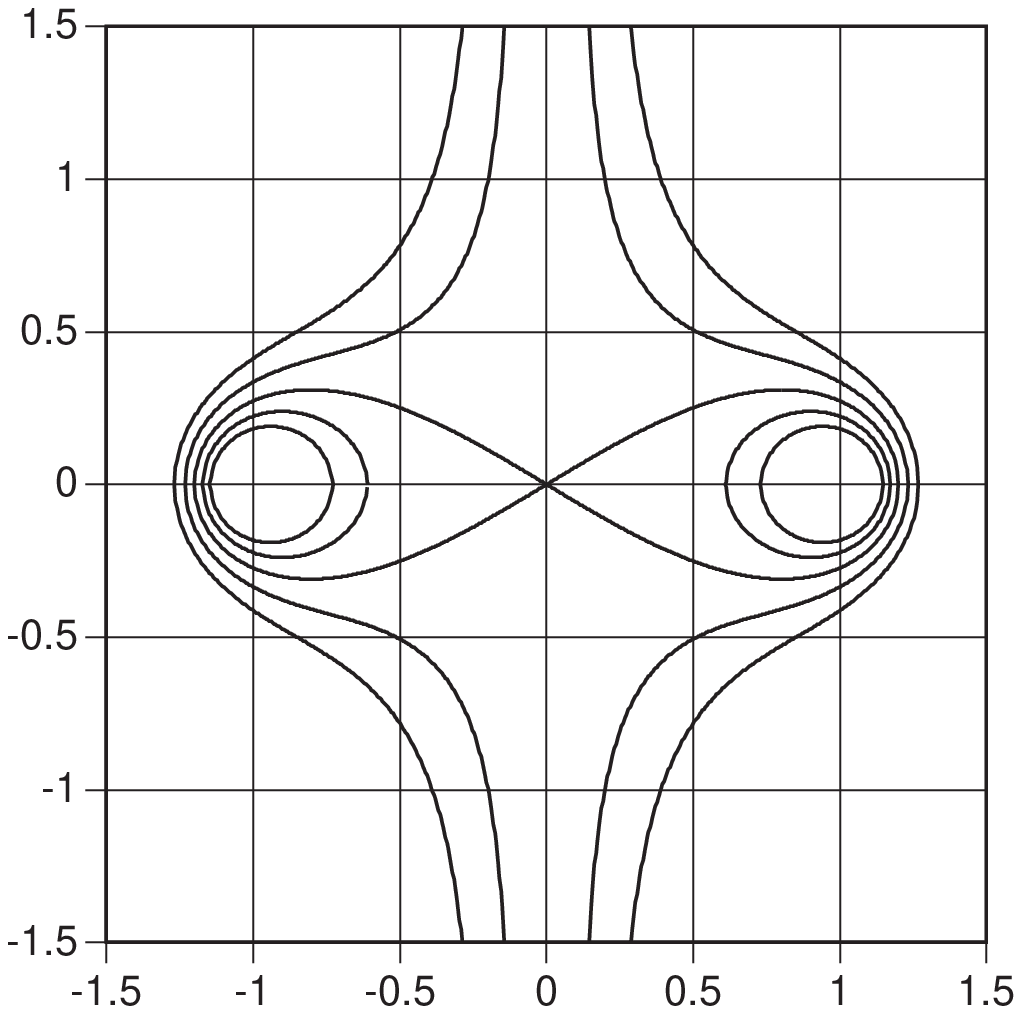}}}} \leftskip 2pc
\rightskip 2pc\noindent{\ninepoint\sl \baselineskip=8pt {\bf Fig.~1}:
The shaded region shows the section of the $w$-plane that maps onto
$Re(z) >0$.  The second diagram shows the $z$-plane, and the curves
correspond to straight lines parallel to the imaginary $w$ axis.  Note
the two lobes that make up the geodesic horizons and the closed orbits
inside these horizons.}
\endinsert

Finally, there is a useful physical model of the BPS geodesics that is
valid for any meromorphic map $w(z)$.  The real and imaginary parts of
$w$ are harmonic functions of $z$, and so straight lines in the
$w$-plane can be thought of as equipotentials of some two-dimensional
field in the $z$-plane.  In models with logarithmic singularities,
this perspective is perhaps most useful when we look at lines that
select the real part of the logarithm.  The singularities can then be
thought of as charges.  In the example above, straight lines parallel
to the imaginary $w$-axis can be thought of as equipotentials of a
uniform electric field in the $x$-direction that has been perturbed by
equal and opposite charges at $z=1$ and $z=-1$ (see Fig.~1).

\newsec{$N=2$, $SU(2)$ Yang-Mills theory with adjoint matter}

\subsec{The curve and differential}

In the formulation of \refs{\RDEW,\HIAM,\EMa}, the Riemann surface and
differential for $N=2$, $SU(2)$ Yang-Mills theory with adjoint matter
can be defined as follows.  One starts with the Weierstra\ss\ torus,
\eqn\btorus{\tilde y^2 ~=~ 4\, (\tilde x - e_1) (\tilde x - e_2)
(\tilde x - e_3) \ ,  }
and constructs a genus two double cover via
\eqn\doubcov{\tilde t^2 - u ~=~ m^2~\tilde x \ .}
The Weierstra\ss\ torus can be uniformized in the familiar manner by
taking $\tilde x = \wp(\tilde \xi)$, $\tilde y = \wp'(\tilde \xi)$.
To avoid confusion later, we will denote the modular parameter of this
torus by $\tilde\tau$.  The differential is
\eqn\lamSWa{\lsw ~=~ \tilde t~d \tilde \xi ~=~ \tilde t~{d \tilde x
\over \tilde y}\ .}
The genus two curve defined by \doubcov, and differential \lamSWa\
have an involution symmetry $\tilde y \to -\tilde y, \tilde t \to
-\tilde t$, and the torus of the $SU(2)$ effective action is obtained
by dividing out this symmetry.  That is, one takes $\tilde z = \tilde
t \tilde y/m$, and replaces $\tilde t^2$ using \doubcov, to obtain
\eqn\doubcov{\tilde z^2 ~=~ (\tilde x + u/m^2)(\tilde x - e_1)
(\tilde x -  e_2) (\tilde x - e_3) \ .}
One can map this to the cubic form of \SW\ by using a fractional
linear transformation \RDEW, however we will map it to different cubic
form that is better adapted to the study of the $m \to \infty$ limit.
To this end, introduce $x = m^2 (\tilde x + b)/(\tilde x + u/m^2)$ and
$y = (x-m^2)^2 \tilde z/m$ where $b = {1 \over 3 e_1} (e_1^2 + 2 e_2
e_3)$.  After some judicious rescaling of $x$ and $y$ the curve and
differential reduce to the form:
\eqn\niceform{y^2 ~=~ (x - \mu^2)(x^2 - \Lambda^4) \ ,  \qquad
\lsw ~=~ c_0~{(x - \mu^2) \over (x - m^2) } ~ {dx \over y} \
, }
where
\eqn\params{\Lambda^2 ~=~ {m^2 \over 3 e_1}~(e_2 - e_3) \ ,
\qquad \mu^2 ~=~ m^2~{u - m^2 b \over u + m^2 e_1} \ , \qquad
c_0 ~=~ m~\sqrt{{m^4 - \Lambda^4 \over m^2 - \mu^2}} \ .}
The whole point of this formulation is that the curve is exactly that
of the pure gauge theory.  Indeed to get this limit one takes $m \to
\infty$ along with $\tilde \tau \to i \infty$.  One then has $(e_2 -
e_3)/ e_1 \to 24\, {\rm e}^{2 \pi i \tilde \tau }$, and one takes
the double limit so that $\Lambda$ remains finite.  One must also make
the infinite additive renormalization: $u \to u + m^2 b$, in order
that the parameter $\mu$ remain finite.  We now take $\Lambda$, $\mu$
and $m$ as the fundamental parameters of the theory, and we will let
$\tau$ denote the Teichm\"uller parameter of the torus \niceform.  One
can easily verify that $\lsw$ does indeed have the property that if
one differentiates it with respect to $\mu$ one gets the holomorphic
differential.

The differential, $\lsw$, has two simple poles with residues $\pm m$,
and a double zero at $x = \mu^2$.  In the pure gauge limit the two
poles coalesce into a double pole.  In the massless, $N=4$ limit ($m^2
\to \mu^2$) these two poles move onto the double zero and all three
annihilate each other.  The finite mass theory thus looks like a
lattice repetition of the toy example above.  Before proceeding to a
more detailed analysis of the horizons we wish to give some rather
useful explicit formula for $w$ as a function of $z$.

\subsec{Integrating $\lsw$}

One can evaluate the indefinite integral in \inteqn, and to do this we
go to the isogenous double cover and uniformize it.  That is, set $x =
\mu^2 + (\Lambda^2 + \mu^2) ~t^2$, and then take \hbox{$t = \cn(\xi,k) 
/ \sn(\xi,k)$}, where $\sn$ and $\cn$ are Jacobi elliptic functions,
and $\xi$ is the flat coordinate of the torus.  One then finds that
\eqn\uniflam{\lsw ~=~ - c~ {(1 ~-~ \sn^2(\xi,k))~ d \xi \over
1 ~-~ k^2 ~\sn^2(\a,k)~ \sn^2(\xi,k)} \ , }
where
\eqn\newparams{\eqalign{k^2 ~&\equiv~ {\theta_2(0|\tau)^4 \over
\theta_3(0|\tau)^4} ~=~ {2 \Lambda^2 \over \Lambda^2 + \mu^2}\ ,
\qquad \sn^2(\a,k) ~=~ {\Lambda^2 + m^2 \over 2 \Lambda^2} \ , \cr
c ~&=~ 2m~\sqrt{{m^4 - \Lambda^4 \over (m^2 - \mu^2)(\Lambda^2 +
\mu^2) }} ~=~ 2 m k^2~{\sn(\a,k)~\cn(\a,k) \over \dn(\a,k)}\ .}}
For later convenience, we note that in the pure gauge limit, the
corresponding result is:
\eqn\uniflampg{\lsw ~=~ 2 \sqrt{\Lambda^2 + \mu^2}~
{\cn^2(\xi,k) \over \sn^2(\xi,k)} ~ d \xi \ .}

The indefinite integrals of these differentials are elliptic integrals
of the third and second kinds respectively.  With a little massaging
the integral of \uniflam\ can be reduced to
\eqn\ansone{w ~=~ w_0 ~+~ (2m~\Ze(\a) - c)~\xi ~+~ m ~\log \bigg(
{\Theta(\xi - \a) \over \Theta(\xi +\a)}\bigg) \ ,}
where $w_0$ is a constant of integration and, following the notation
of \WW,
\eqn\ellfns{\Theta(\xi) ~\equiv~ \theta_4\bigg({\xi \over
\theta_3(0|\tau)^2} \bigg | \tau \bigg) \ ; \qquad
\Ze(\xi) ~=~ {\Theta'(\xi ) \over \Theta(\xi)} \ .}
For the pure gauge theory one gets
\eqn\anstwo{w ~=~ w_0 ~-~ {2 \over K} ~\sqrt{\Lambda^2 + \mu^2} ~
\bigg(~E~\xi ~+~{\pi \over 2} ~{\theta_1'(\pi \xi/2 K) \over
\theta_1(\pi \xi/2 K)}~\bigg) \ ,}
where, following the usual convention $E$ and $K$ are elliptic
periods.  Explicitly, one has:
\eqn\ellperiods{K ~\equiv~ {\pi \over 2}~ \theta_3(0|\tau)^2 \ ,
\qquad E ~\equiv~{1 \over 3}~ \Big(2 - k^2 - \theta_1'''(0|\tau)/
(\theta_1'(0|\tau)~\theta_3(0|\tau)^4) \Big)~K \ .}
Expressions for the integral of $\lsw$ in the pure gauge theory were
also given in \refs{\ABFF,\FFer}.

{}From \ansone\ it is trivial to read off the periods $a$ and $a_D$.
One gets
\eqn\aandaDone{\eqalign{a ~&=~ 2(2m~\Ze(\a) - c)~K + 2 \pi i n_1 m
\ , \cr \qquad a_D ~&=~  2(2m~\Ze(\a) - c)~iK' ~+~ 2\pi i m(\a/K +
n_2 ) \ ,}}
where $K' = - i \tau K$, and $n_1$ and $n_2$ are the winding numbers
around the simple poles at $\xi = i K' \pm \a$.  Similarly, for
\anstwo\ one gets
\eqn\aandaDtwo{a ~=~ 4~\sqrt{\Lambda^2 + \mu^2}~E \ , \qquad a_D ~=~
2~\sqrt{\Lambda^2 + \mu^2}~\Big(2 \tau E - {\pi \over K}\Big)\ .}

These expressions enable one to analytically solve for the geodesics
without numerical integration.  In particular, one can find all the
geodesics through some point, $\xi_0$, by plotting all the curves on
which $\arg(w - w(\xi_0))$ is constant.  This approach was used to
create Figs.~2 and~3.

\goodbreak\midinsert
\centerline{\epsfxsize 3.truein\epsfbox{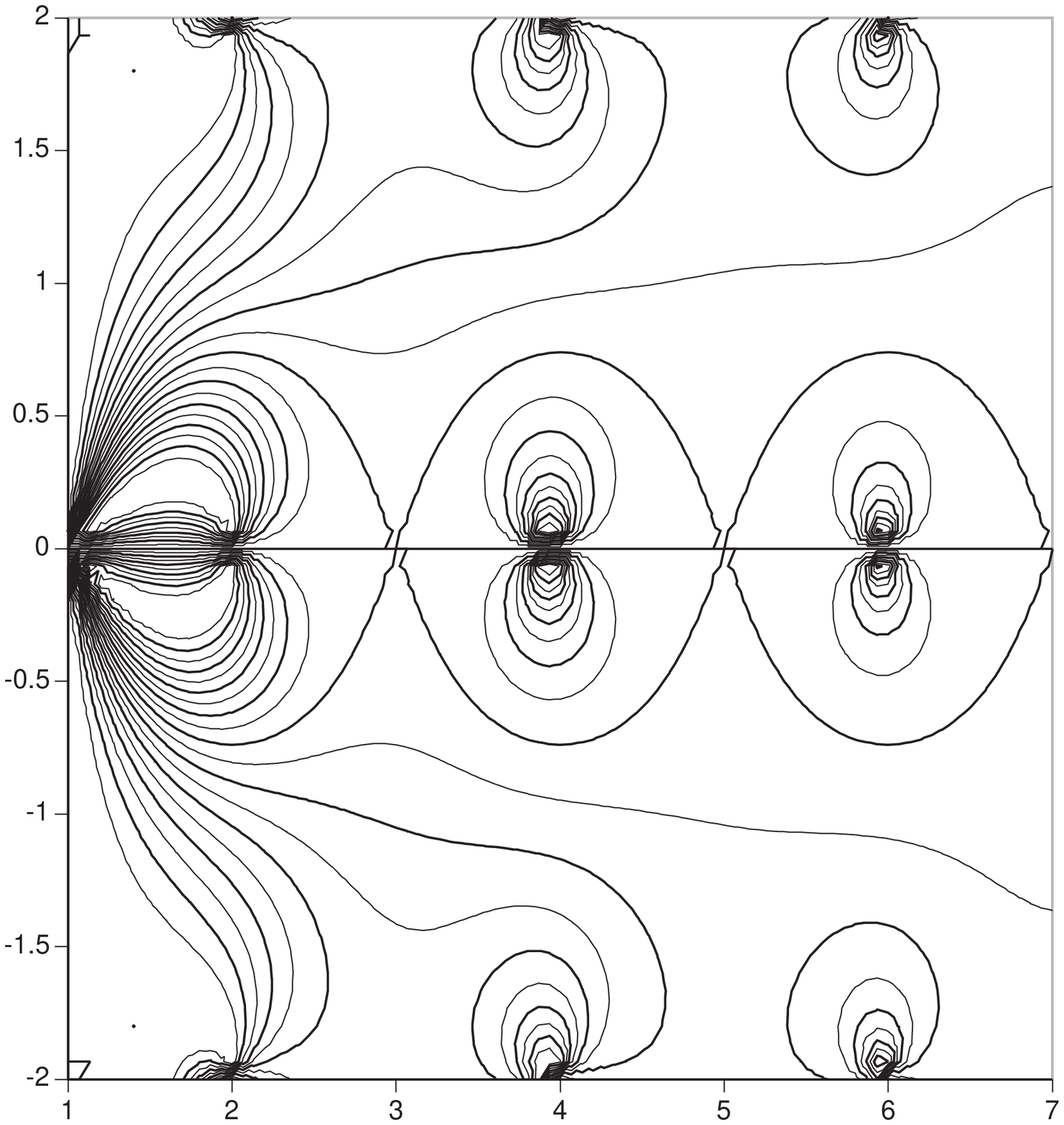}}\leftskip 2pc
\rightskip 2pc\noindent{\ninepoint\sl \baselineskip=8pt {\bf Fig.~2}:
The lines shown are geodesics drawn in the $z$-plane.  The horizontal
and vertical axes are marked off in units of $K$ and $K'$ (the
half-periods), and we have taken $\tau = i$, or $K = K'$.  There are
zeroes of $\lsw$ at the points $(2 p + 1,2 q )$, and double poles at
the points $(2p, 2q)$, $p, q \in \ZZ$.  The horizons around the poles
at $(2n,0)$ are evident, and are comprised of geodesics corresponding
to $W$-bosons.}
\endinsert

\newsec{Stability of BPS states in $SU(2)$ Yang-Mills theory}

We now wish to show how to obtain the BPS spectrum of $N=2$, $SU(2)$
Yang-Mills theory, and of the same model coupled to an adjoint
hypermultiplet, by finding the geodesic horizons.  To do this one
simply needs to look for all geodesics that start and finish at zeroes
and surround poles of $\lsw$.  We start by showing how the geodesic
method easily replicates the known results for the pure gauge theory
\refs{\SW,\ABilal,\ABFF}.

\subsec{The pure gauge theory}

The Seiberg-Witten differential has a double pole (with vanishing
residue) and a double zero.  In the uniformization used in the last
section, they are located at $\xi = K$ and $\xi = 0$ (mod $2K$ and
$2iK'$) respectively.  The geodesic horizon is thus reducible, its
irreducible parts must correspond to a collection of BPS states.  For
large $\mu$, the $W$-boson is the BPS state of lowest mass, and so the
geodesic horizon for large $\mu$ must be the pair of geodesics
corresponding to a $W^+$ and $W^-$ connecting two zeroes above and
below the double pole.  This fact is born out by the manifest horizons
in Fig.~2.  These horizons mean that any BPS state can only cross the
$W$-boson trajectory at the zero of $\lsw$.  It follows immediately
that any BPS state of magnetic charge larger than one must be
reducible to BPS states of magnetic charge one.  Thus the existence of
the stable $W$-boson implies that only the $W$-boson itself, and
states with magnetic charge one can be stable\foot{The instability of
states of charge $(n,0)$ and $(0,n)$ is obvious in the geodesic
approach for the reasons outlined in \KLMVW.}.

At the curve of marginal stability one knows that the $W^\pm$-bosons
become unstable to decay into states of monopole and electric charges
$\pm (1,0)$ and $\pm (1, -1)$.  This is very clearly seen in even the
most rudimentary numerical calculation of the geodesic horizons (see,
for example, Fig.~3).  As $\tau$ approaches the curve of marginal
stability, the $W^\pm$-boson geodesic horizons deform to the zeroes of
$\lsw$ located at $\xi = K \pm 2iK'$ (mod $2K$ and $2K'$). At and
beyond the curve of marginal stability, the geodesic horizon around
the pole is the quadrilateral with edges $(\pm 1,0)$ and $(\mp 1, \pm
1)$.  These quadrilateral horizons in every fundamental region of the
torus completely block passage of BPS states in every direction: To
get a winding number of more than $\pm 1$ in either the $(1,0)$ or
$(-1,1)$ directions, a~BPS state can only pass through the zeroes at
the quadrilateral corners, and thus must be decomposable.  Thus the
fact that the $W$ boson is unstable to a monopole-dyon pair inside the
curve of marginal stability implies that the monopole and dyon are the
{\it only} stable BPS states.

\goodbreak\midinsert
\centerline{{\epsfxsize 5cm \epsfbox{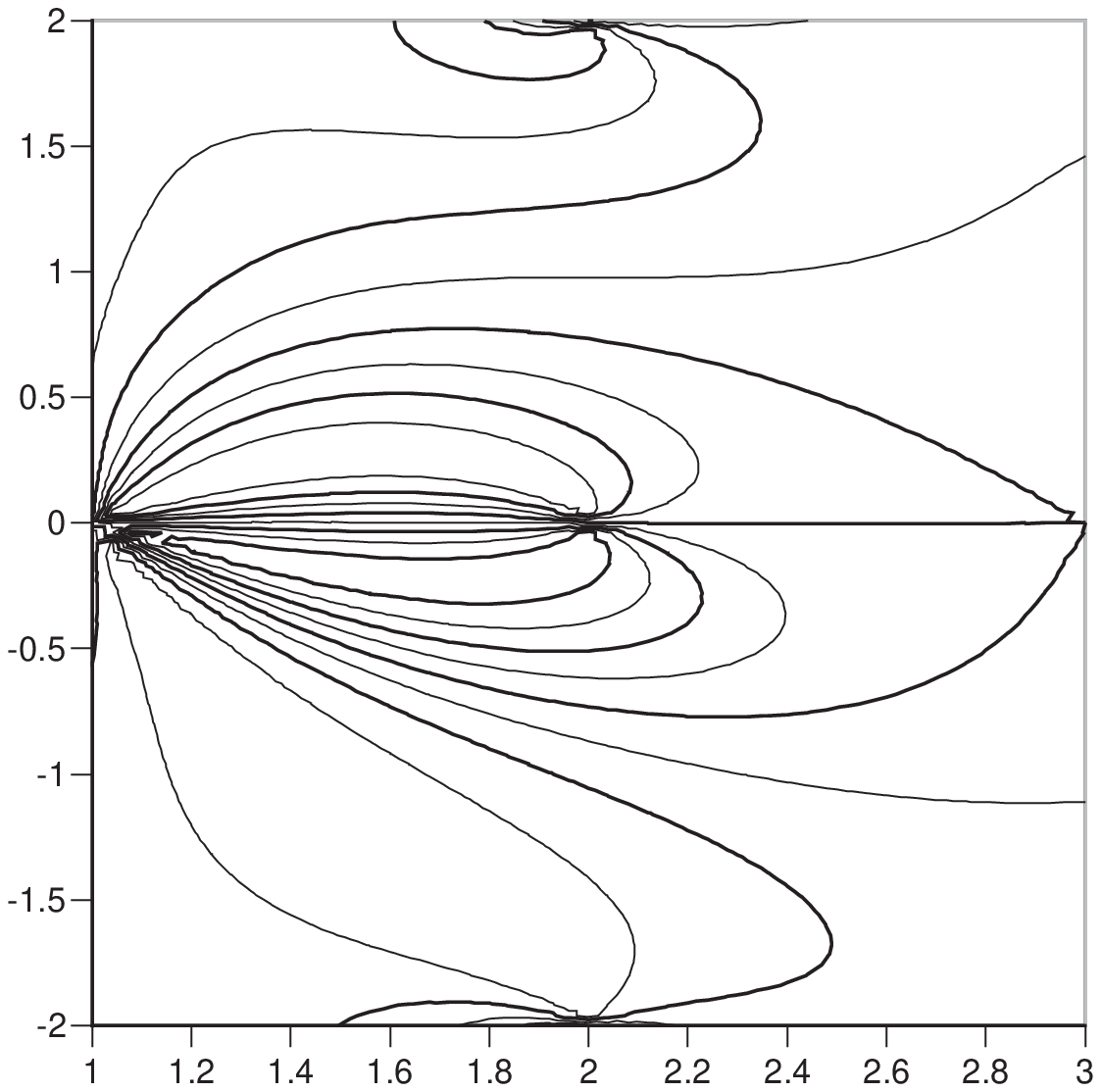}} \hskip 2cm
{\epsfxsize 5cm \epsfbox{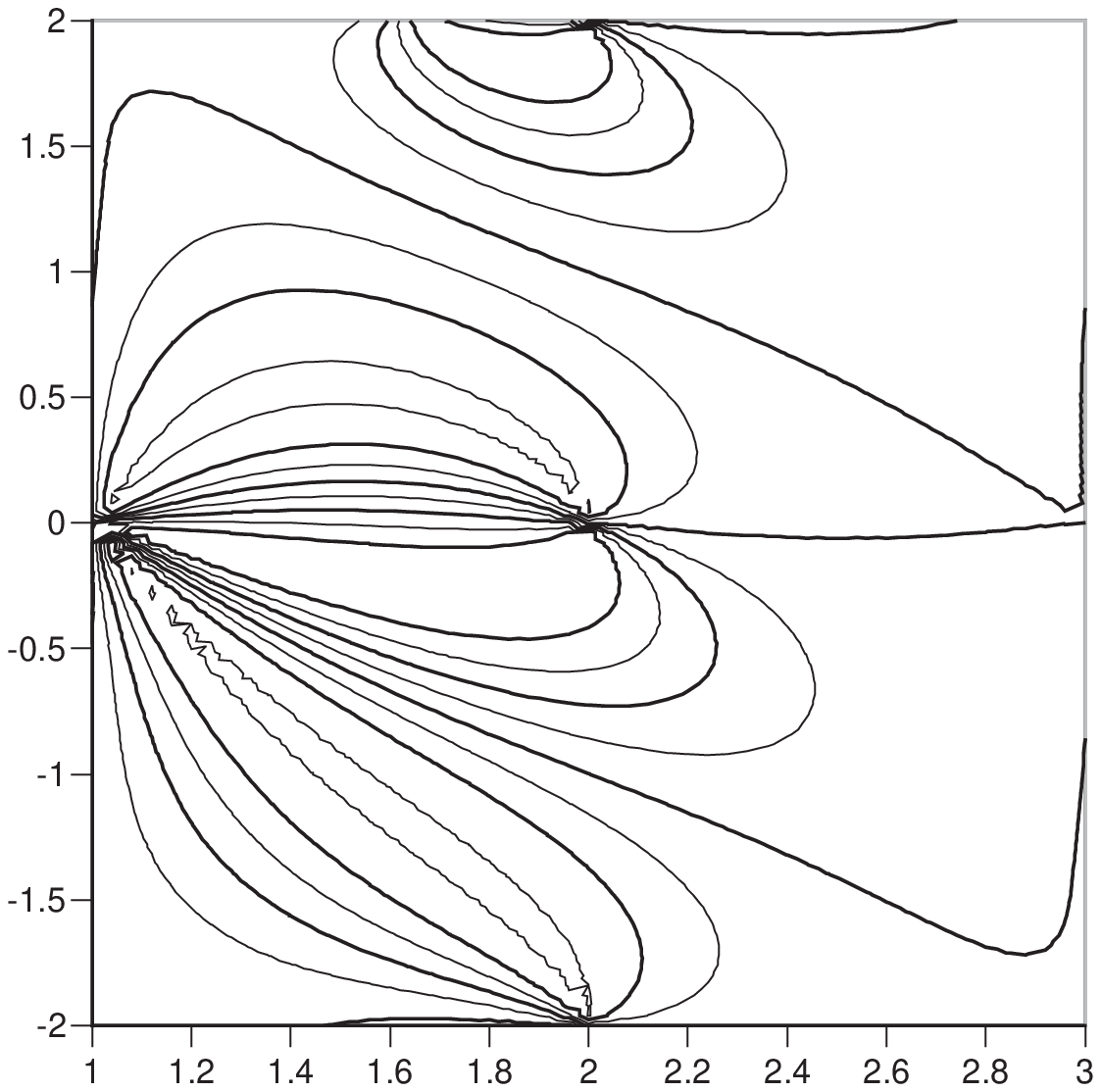}}}\leftskip 2pc
\rightskip 2pc\noindent{\ninepoint\sl \baselineskip=8pt {\bf Fig.~3}:
These two diagrams show several geodesics and a single geodesic
horizon around a double pole at $(0,2)$.  The first diagram has $\tau
= 0.3 + 1.0 i$, and the second has $\tau = 0.3 + 0.654 i$.  The curve
of marginal stability passes through the point $\tau = 0.3 + 0.651 i$.
As one approaches the curve of marginal stability, the geodesic
horizon deforms to the zeroes at $(1,2)$ and $(3,-2)$.  On and inside
the curve of marginal stability the geodesic horizon becomes the
quadrilateral with vertices $(1,0), (1,2), (3,0)$ and $(3,-2)$.}
\endinsert

One can, in fact, infer the entire structure of the BPS spectrum
without doing any computations or simulations.  {}From the diagrams in
\KLMVW\ one can immediately see that there is a curve (homotopic to a
circle) in the parameter space on which the $W$-boson is marginally
stable to a monopole and a dyon.  Thus the geodesic horizon must
deform and touch another zero of $\lsw$ as described above.  The fact
that the quadrilateral horizon persists inside the curve of marginal
stability is then inferred from the monodromies of the theory.
Specifically, one considers what happens as $\tau$ of the torus is
decreased from $i \infty$ to near the real axis.  When $Im(\tau)$ is
decreased far enough, one must get to a modular inversion of the
strong coupling region again.  This means that the geodesic horizons
must skip from surrounding $a$-cycles of the torus to surrounding some
combination of $a$ and $b$-cycles.  Such horizon jumps can only occur
at curves of marginal stability, and the only way that it can happen
so that the monodromies of the theory is respected is if the
quadrilateral forms and persists inside the curve of marginal
stability, and then it detaches to leave a horizon along the $a$ or
$b$ cycle depending on where one crosses the curve marginal stability
considered as a function of $\tau$.

The beauty of the foregoing argument is that it only cares about the
modular properties of the theory, and how the zeroes and poles (the
divisor) of $\lsw$ moves as a function of the parameters.

\subsec{The softly broken $N=4$ theory}

This differential $\lsw$ now has three independent periods, and as we
will see, the BPS spectrum is richer, but computable.

Consider first the unbroken $N=4$ theory.  As can be seen from
\niceform, the differential $\lsw$ collapses to the holomorphic
differential of the torus:
\eqn\refa{
   y^2 = x\,(x^2-\Lambda^4)\ ,
   \ \ \ \lambda_{{\rm SW}}~=~ c_0~{dx \over y} ~=~ c_0 d \xi\ .}
The geodesics are thus straight lines on the torus itself, and so the
spectrum consists of any dyon $(g,q)$ with $g$ and $q$ relatively
prime integers \refs{\ASen,\MPor}.

Turning on the mass, $m$, of the hypermultiplet causes the
differential, $\lsw$, to develop a double zero and two simple poles
with residues $\pm m$.  The geodesic horizons near the zeroes and
poles look like those of the toy model -- two lobes around the poles
with the lobes touching at the double zero.  The conformal mapping,
$w(z)$, given by \ansone\ is thus relatively easy to describe: The
$w$-plane is broken into fundamental regions of periods $a$ and $a_D$.
Inside each such region there is a cut parallel to the $Im(w)$-axis,
and of length $2 \pi m$.  As in the toy example, the cut in the
$w$-plane is doubly covered by the two geodesic horizons in the
$z$-plane, and to the left and right of the $w$-plane cuts are
semi-infinite cylinders.  If one does not cross these cuts the mapping
$z(w)$ is single valued.  Any straight line through a cut must pass
into a cylinder, and so viable BPS geodesics are straight lines that
avoid all the cuts.

The cuts or geodesic horizons affect BPS geodesics in two different
ways.  First, they confine the geodesics to either terminate on a
zero, or ``squeeze through'' the ``passes'' between two such horizons.
Secondly, they mean that BPS states can only pass around the pole on
one side and not on the other: the horizon thus imposes restrictions
on the hypermultiplet charges that a BPS geodesics can pick-up.  To be
more specific, given a BPS geodesic, we can always compare its
hypermultiplet charge to that of a fixed reference state such as a
composite of monopoles and $W$-bosons (or $(1,1)$ -dyons).  The BPS
geodesic must either wind around {\it both members} of a pair of
simple poles, thus collecting a net zero contribution to its
hypermultiplet charge, or it must terminate at the zero to which the
poles' geodesic horizons connect.  From the perspective of the
$z$-plane there are apparently sixteen ways in which one can connect
two zeroes, but only four of them will be straight lines in the
$w$-plane (see, for example, Fig.~4).  One can thus easily see that
BPS states with fixed magnetic and electric charges can come in
``multiplets'', with at most most four members, and with with
hypermultiplet charges of $0, \pm m$.  For example, Fig.~4 shows the
possible BPS states with electric charge $+1$ and magnetic charge $0$.
These four states have a very simple perturbative interpretation.  We
are breaking $N=4$, $SU(2)$ Yang-Mills to an $N=2$, $U(1)$ gauge
theory, and so the charge $+1$ BPS states should not only include the
$W^+$, $N=2$ vector multiplet, but also should contain a charge $+1$,
complex hypermultiplet, which contains states with opposite
hypermultiplet charge.  The four contours represent two copies of the
$W^+$ boson and the pair of charged hypermultiplet states\foot{The
geodesic horizon itself can be thought of as the neutral
hypermultiplet state.}.  Since the $N=4$ theory is $SL(2,\ZZ)$
invariant, we can work perturbatively around any winding number state,
and reach the same conclusion.  Thus we learn that even at strong
coupling, and for any value of the mass $m$, there are no new BPS
states other than the ones that one would expect from perturbation
theory around all the phases of the softly broken $N=4$ theory.  Our
purpose now is to show the extent to which the $N=4$ BPS states
destabilize as the mass is increased.

\goodbreak\midinsert
\vskip .5cm
\centerline{  {\epsfxsize 5cm \epsfbox{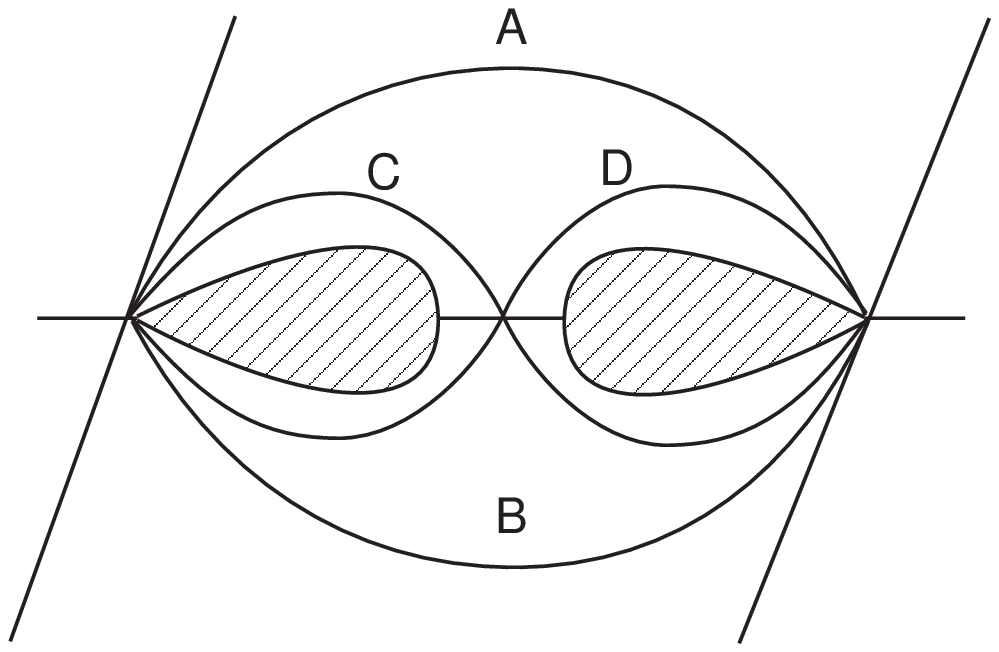}}
\hskip 1.5cm {\epsfxsize 5cm \epsfbox{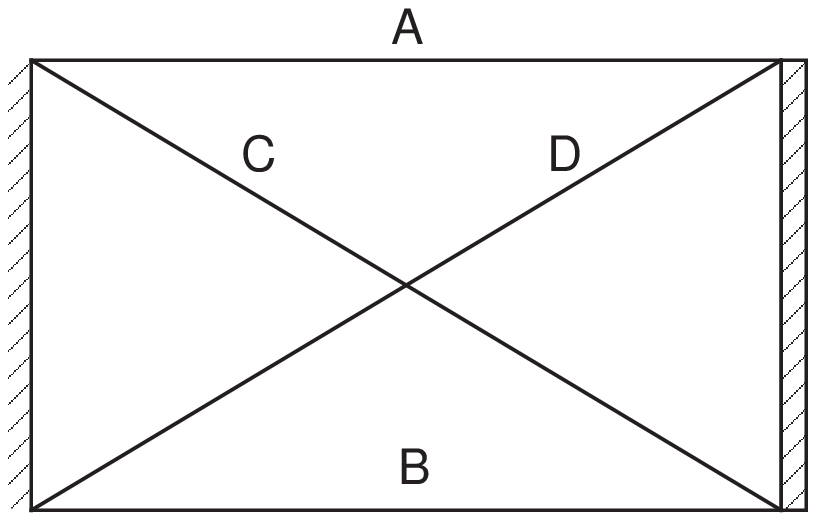}} } \leftskip 2pc
\rightskip 2pc\noindent{\ninepoint\sl \baselineskip=8pt {\bf Fig.~4}:
The irreducible geodesics with $(g,q) = (0, \pm 1)$.  The first figure
shows them in the $z$-plane and the second shows them in the
$w$-plane.  The shaded regions in the $z$-plane are the interiors of
the horizons, and in the shaded lines in the $w$-plane are branch
cuts.}
\endinsert

If one approaches the pure gauge limit, then the simple poles in the
$z$-plane approach one another at a point a half-period away from the
zeroes.  The geodesic horizons of the two approaching simple poles
flatten at the ends, but cannot touch since they are irreducible
geodesics.  Thus until the two simple poles actually coincide, there
is always a ``hypermultiplet pass'' between these horizons.  The
distance through this path gets infinitely long as $m \to \infty$.
Put another way, the two $W$-boson geodesics, labeled $A$ and $B$ in
Fig.~4, form a selective horizon in the sense of section 2 -- one can
recross it by weaving in between the simple poles that it encloses.

The corresponding picture in the $w$-plane is as follows: The
$w$-plane has cuts that periodically repeat with periods $a$ and
$a_D$, and the length of the cuts is proportional to $m$.  Outside the
curve of marginal stability of the $W$-boson, the cuts fall into
well-defined rows that do not interleave, creating an open channel
between one row of cuts and the next.  The edges of these channels are
defined by the periodic repetition of the $W$-boson geodesic (see
Fig.~5).  As $m$ gets large the only fundamental BPS states that have
small mass correspond to those irreducible geodesics that remain
within the small channel between the tops of one row of cuts and the
bottoms of the immediately neighbouring row of cuts.  These states
have charges $(g,q)$ equal to $(0,n)$ or $(\pm 1,n)$, and become the
stable states of the pure gauge limit.  A chasm of width $m$ opens out
under the other states.  However the identification in the $z$-plane
of the endpoints of the cuts provides bridges of length zero across
the chasm.  The cost of using these bridges is the loss of
irreducibility for the BPS state.  Inside the curve of marginal
stability of the $W$-boson, the rows of cuts interleave.

We can refine the $w$-plane picture to give much more specific
information about individual states.  Choose some cut $C_0$ as a base
``point''.  There are then stable BPS states associated with another
cut, $C_1$, if and only if one can draw straight lines between the
end(s) of $C_1$ and the end(s) of $C_0$ in such a manner that these
straight lines do not meet any other cuts in between.  Suppose that
there is indeed a full complement of four BPS states associated with
$C_1$.  Draw the straight lines on the $w$-plane associated with these
states and {\it extend} these lines beyond $C_1$ (see Fig.~5).  Think
of these extended lines as the edges of a {\it shadow} cast by $C_1$.
That is, think of $C_0$ as an extended light source, and $C_1$ as a
casting a shadow.  Note that the light source and the shadow casting
object have the same size.  The lines that we have drawn define the
edges of the umbra and penumbra of the shadow.  If a point $w$ lies in
the umbra of the shadow of $C_1$ then no straight line can connect it
to $C_0$ without passing through $C_1$.  If $w$ lies in the penumbra
of the shadow, then a straight line (that does not cross $C_1$) can
connect $w$ to either the top or bottom of $C_0$ but not to both.
Therefore the extent to which there are stable BPS states of charge
$(g,q)$ can be determined by the extent to which the cut, $C$, located
at $g a_D + q a$ relative to $C_0$, lies in the shadows of the cuts
between $C$ and $C_1$.  We can, of course, transfer these cuts and
shadows to the $z$-plane and think the of shadows as being cast by
geodesic horizons.

Thus, given the mass, $m$ and the Higgs vev, one can use the formulae
of section 3 to compute $a$ and $a_D$, and it then becomes an
elementary geometric exercise to determine if there is a stable BPS
states with a given set of charges.

\goodbreak\midinsert
\centerline{{\epsfxsize 8cm \epsfbox{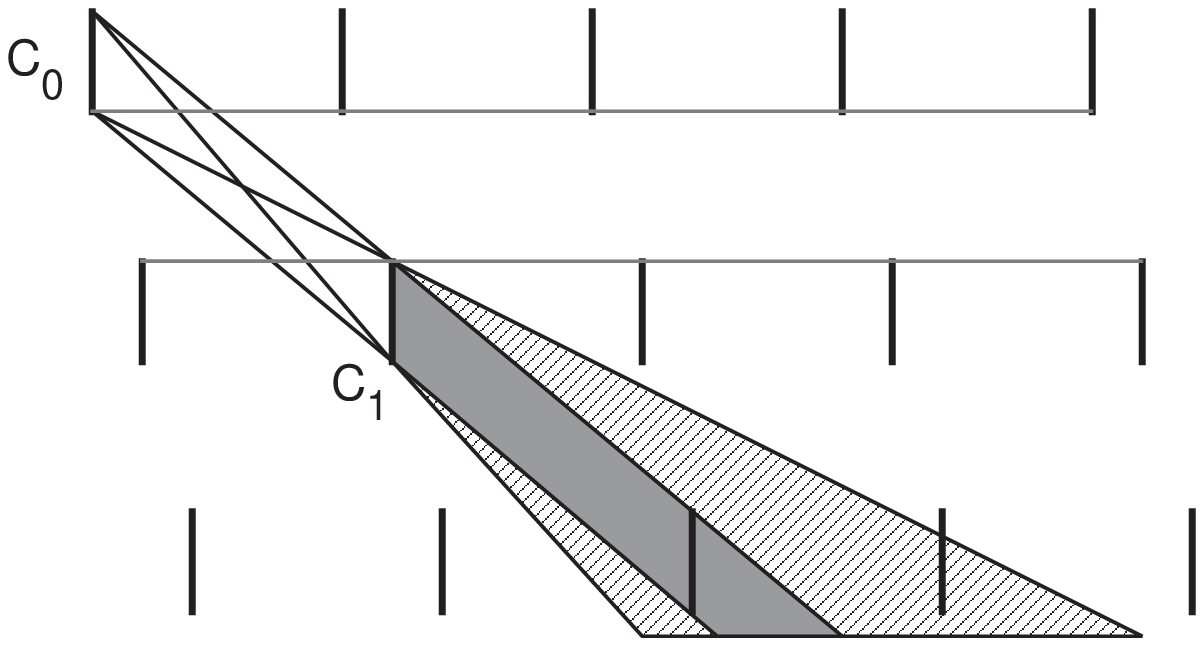}}} \leftskip 2pc
\rightskip 2pc\noindent{\ninepoint\sl \baselineskip=8pt {\bf Fig.~5}:
The $w$-plane of the model with adjoint matter.  The heavy vertical
lines are branch cuts whose length is proportional to the mass m. The
horizontal dotted lines consist of $W$-boson geodesics, and define the
sides of a channel between rows of cuts.  The shaded regions show the
umbra and penumbra of a shadow cast by the cut $C_1$.}
\endinsert

The idea of shadows can be thought of as a natural geometric extension
of the fact that stable BPS states in the $N=4$ theory must have
$(g,q)$ relatively prime: here the shadows are the shadows cast by the
points $(g/h, q/h)$ where $h$ is the highest common divisor of $g$ and
$q$.  As one turns on the mass, the shadows get fatter. There are some
obvious general statements that can be made.  First, if $g$ and $q$
are not relatively prime then the entire corresponding cut lies in the
full umbra of the cut at $(g/h, q/h)$, and so {\it all} BPS states
with magnetic and electric charges $(g,q)$ are unstable - they cannot
be stabilized by picking up hypermultiplet charge.

More generally, as $g$ and $q$ (now taken to be relatively prime) get
larger and larger, there are potentially more shadows that can fall on
them, and so as $m$ increases, the states with larger $g$ and $q$ will
generically destabilize first.  Since the width of an umbra does not
change as one moves along it, but the width of the penumbra increases
with distance, one will also generically find that some components of
a geodesic multiplet will destabilize long before the other components
of the same multiplet destabilize.  Indeed it is rather easy to
convince oneself that for fixed $m$, there are always suitably large
values of $g$ and $q$ beyond which the corresponding cuts are always
in some penumbra\foot{The set of interiors of all penumbras form an an
open cover of any closed, finite angular interval not including the
two directions in which the cut $C_0$ points.  There is thus always a
finite subcover of such an interval.}.  However, the same is not true
of an umbra since the opening angle of the umbra is zero.  Indeed, if
one is outside the curve of marginal stability of the $W$-boson, one
can check that there is always one state with charges $(g,q) =
(-n,n+1)$ for any $n$: the top of $C_0$ can be always be connected via
a straight line to the bottom of the cut corresponding to charges
$(-n,n+1)$.  One can also easily see that in this region of moduli
space there is a full geodesic multiplet of states with charges
$(-n,1)$, but these have mass larger than $m$ and hence decouple in
the pure gauge limit.

There are also always some states of charge $(\pm 1,n)$, and when $n$
small enough compared to $|a_D/m|$ there will always be a full
multiplet of geodesics\foot{In defining $a_D$ we allowed arbitrary
additions of $2 \pi i m$.  This is indexed by $n_2$ in \aandaDone.
Here we choose the value of $n_2$ that gives the smallest value to
$|a_D|$.}.  As $n$ increases, some of the geodesics are excluded by
shadows.  For large $n$ (or large $m$) there is only one geodesic left
in the multiplet, and it is one of the geodesics that was previously
identified as representing an $N=2$ hypermultiplet.  This
identification is nicely consistent with the pure gauge limit in which
these states become the dyons.

\subsec{Curves of marginal stability}

We wish to conclude this section by making one or two comments related
to curves of marginal stability for the massive model.  Our intent is
not to make an exhaustive study here, but draw attention to some
interesting features.  To get such curves, one first imagines fixing
one of the two scale independent complex moduli, $\alpha$ and $\tau$,
and then varying the other.  One then chooses a particular physical
state, and then one seeks curves (in the space of the variable
modulus) on which that state becomes unstable.  There are, of course,
infinitely many possibilities for the massive model, but they are
easily characterized: One simply draws the state of interest on the
$w$-plane (a straight line on Fig.~5), and varies the parameters until
that state touches the top or bottom of some intervening cut, at which
point the physical state of interest becomes a composite (reducible)
state.

As mentioned above, the $W$-boson's curve of marginal stability is
heralded by the interleaving of the rows of cuts in Fig.~5.  It is
also interesting to note that if one crosses the curve of marginal
stability of the $W$-boson then even though it decays into a monopole
and a dyon, the other (hypermultiplet) geodesics with the same
magnetic and electric charge as the $W^\pm$ remain stable until the
interleaving of cuts disturbs them, too.

Finally we turn to the curves of marginal stability for the geodesic
horizons (neutral hypermultiplets) themselves.  This is relevant to
understanding what happens to the horizons and BPS states as one
smoothly changes the position of the poles of $\lsw$.  That is, one
smoothly changes $\alpha$ in \ansone\ by any period $2 K$ or $2 i K'$
of the fundamental region in the $z$-plane.  This must deform the
horizons in the direction in which $\alpha$ moves, but, if one changes
$\alpha$ by a period, the horizons must return to their original
configuration.  This means that the horizon must shift its attachment
from one zero of $\lsw$ in $\widetilde \Sigma$ to another zero.  It
does this by momentarily becoming reducible and connecting to two
zeroes of $\lsw$ that are separated by a period.  There are thus lines
of marginal stability for the horizons.  If one changes $\alpha$ by $2
K$ then the line of marginal stability is a subset\foot{There are
other states that destabilize on other parts of the set $Im(a/(2\pi
im)) = 0$.}  of the curve $Im(a/(2\pi im)) = 0$, and similarly if one
changes $\alpha$ by $2 K'$ then the line of marginal stability is
contained in the curve $Im(a_D/(2 \pi i m)) = 0$.  On either side of
these lines of marginal stability the geodesic horizon detaches from
one zero or the other and becomes irreducible again.

It is amusing to note that if one is on such curves of marginal
stability, then the geodesic horizons generate a wall much like that
of the pure gauge theory, and if the wall is in the direction of the
$W$-bosons then all the irreducible BPS states with $|g| > 1$ cease to
be stable.  Unlike the pure gauge theory, this instability of BPS
states is ephemeral -- it only happens {\it on} the curve of marginal
stability, and not in some region to one side of it.  If one is at the
intersection of the curves of marginal stability in both the $K$ and
$K'$ directions (which also means that one is also on the curve of
marginal stability of the $W$-boson), then the geodesic horizon
attaches to three zeroes of $\lsw$, and one momentarily gets something
just like the ``quadrilateral horizon'' of the pure gauge theory, with
similar consequences for the BPS spectrum.

\newsec{Conclusions}

We have shown that the BPS geodesics can be used with relative ease to
give a simple geometric understanding of the stability of BPS states
in $N=2$, $SU(2)$ pure gauge theory and in the theory with adjoint
matter.  The key to the analysis is to find geodesics that partition
the covering space of the Seiberg-Witten Riemann surface.  This led us
to introduce geodesic horizons and the shadows cast by such horizons.
It seems very likely that this approach will find application in other
models.

Since the geodesic method was first introduced in \KLMVW, it was
evident that the poles of $\lsw$ played a crucial role in
destabilizing BPS geodesics.  As can be see from this work, the zeroes
of $\lsw$ are just as important, and the degree of the poles is
crucial to the horizon structure.  Thus we believe that the general
structure of the BPS spectrum is ultimately determined by the full
divisor class of $\lsw$.  For example, if one takes the pure gauge
theory and adds matter in the fundamental representation, then the
divisor class of $\lsw$ changes, even if the matter is massless.  It
is this fact that must be the crucial ingredient in understanding the
appearance of new stable BPS states as the number of flavours is
increased.

Since we have exhibited the implicit general solution to the geodesic
problem, we cannot help but note its remarkable form.  First, the
expression in terms of theta functions leads to natural conjectures
for generalizations to higher genus surfaces.  (Indeed, if true, this
is how the divisor class of $\lsw$ would mathematically determine the
geodesic problem.)  On a far more speculative level, there might be
some interesting topological interpretations of the theta functions.
Recall that the BPS geodesics represent the physical states of a
topologically twisted self-dual string \KLMVW.  The characterization
of these stringy states as ``straight lines in the $w$-plane'' is also
all too reminiscent of the characterization of solitons in massive
$N=2$ supersymmetric field theories in two dimensions \FMVW.  It is
thus tempting to try to interpret $w$ as some kind of effective
superpotential for the massive string.  If $w$ is linear in $\xi$ then
it is {\it a priori} a trivial superpotential.  However, the
topological charge of a soliton is given by $\Delta w$, and so a
linear superpotential is appropriate to a free theory with winding
states.  While it may be a coincidence, it would be interesting to try
to interpret the logarithms of the theta functions as some kind of
effective action (or superpotential) induced by integrating out
bosonic and fermionic degrees of freedom that have been twisted by
some parameter $\alpha$.  These degrees of freedom could conceivably
be found in the other modes of the $3$-brane from which the self-dual
string descends.

\goodbreak
\vskip2.cm\centerline{\bf Acknowledgements}
\noindent
We would like to thank W.~Lerche and J.~Rabin for valuable
discussions.  N.W.~is also grateful to the theory division at CERN for
hospitality where this work was begun.  This work is supported in part
by funds provided by the DOE under grant number DE-FG03-84ER-40168.

\goodbreak
\listrefs
\end